\documentclass[aps,prl,twocolumn,showpacs]{revtex4-1}
\usepackage[T1]{fontenc}
\usepackage{graphicx}
\usepackage{color}
\usepackage{amsmath}
\usepackage{ulem}

\newlength{\halfcolumnwidth}
\begin{document}
\title{Sub-Poissonian fluctuations in 
a 1D Bose gas: from the quantum quasi-condensate to the strongly interacting regime}


\date{\today}
\author{Thibaut Jacqmin$^{(1)}$, Julien Armijo$^{(1)}$, Tarik Berrada$^{(1,2)}$, Karen V. Kheruntsyan$^{(3)}$ and
Isabelle Bouchoule$^{(1)}$}
\affiliation{$^{(1)}$Laboratoire Charles Fabry, Institut
d'Optique, UMR8501 du CNRS, 91127 Palaiseau Cedex, France\\
$^{(2)}$Vienna Center for Quantum Science and Technology, Atominstitut, TU Wien, 1020 Vienna, Austria\\
$^{(3)}$School of Mathematics and Physics, The University of Queensland, Brisbane, Queensland 4072, Australia
}

\begin{abstract}

We report on local, \textit{in situ} measurements of atom number 
fluctuations in 
slices of a one-dimensional Bose gas on an atom chip setup. 
By using current modulation techniques 
to prevent cloud fragmentation, we are able to
probe the crossover from weak to strong interactions. 
For weak interactions, fluctuations go continuously from 
super- to sub-Poissonian as the density is increased,
which is a signature of the transition between the sub-regimes
where the two-body correlation function is dominated 
respectively by thermal and quantum contributions.
At stronger interactions, the super-Poissonian region disappears, and 
the fluctuations go directly from Poissonian to  sub-Poissonian, as 
expected for a `fermionized' gas.
\end{abstract}


\maketitle


Fluctuations witness the
interplay between quantum statistics and interactions and therefore their measurement
constitutes an important probe of 
quantum many-body systems.
In particular, measurement of atom number fluctuations in ultracold 
quantum gases has been a key tool
in the study  of
the 
Mott insulating phase in optical lattices \cite{Folling2005}, isothermal 
compressibility  of Bose and Fermi 
gases \cite{Esteve06,Armijo10_2,Sanner10,Mueller10}, 
magnetic susceptibility of a strongly interacting 
Fermi gas \cite{Sanner2011}, scale invariance of 
a two-dimensional Bose gas \cite{Hung2011}, generation of atomic entanglement in 
double-wells \cite{Esteve2008}, and relative number squeezing in pair-production via binary collisions 
 \cite{Jaskula2010,TwinAtomBeams}.


While
a simple account of quantum statistics can change the atom number
distribution, in a small volume of an ideal gas, from a classical-gas
Poissonian to super-Poissonian (for bosons) or sub-Poissonian (for
fermions) distributions, 
many-body processes
can further modify the correlations and fluctuations.  For
example, three-body losses may lead to sub-Poissonian fluctuations in
a Bose gas~\cite{Whitlock2010,Itah2010}. Even without dissipation, the
intrinsic interatomic interactions can also lead to sub-Poissonian
fluctuations, such as in a repulsive Bose gas in a periodic lattice
potential, where the energetically costly atom number fluctuations are
suppressed.
This effect has been
observed 
for large ratios of
the on-site interaction energy to the inter-site tunnelling energy 
\cite{Wei2007,Gross2010}, with the extreme limit 
corresponding to the Mott insulator phase 
~\cite{Bakr2010,Sherson2010}.
The same physics, 
accounts for sub-Poissonian fluctuations 
observed in double-well experiments~\cite{Sebby-Strabley2007,Esteve2008}.
Sub-Poissonian fluctuations of the total atom number have been also realised
via controlled loading of the atoms into very shallow traps
~\cite{Chuu2005}.

\begin{figure}
\includegraphics[width=7.9cm]{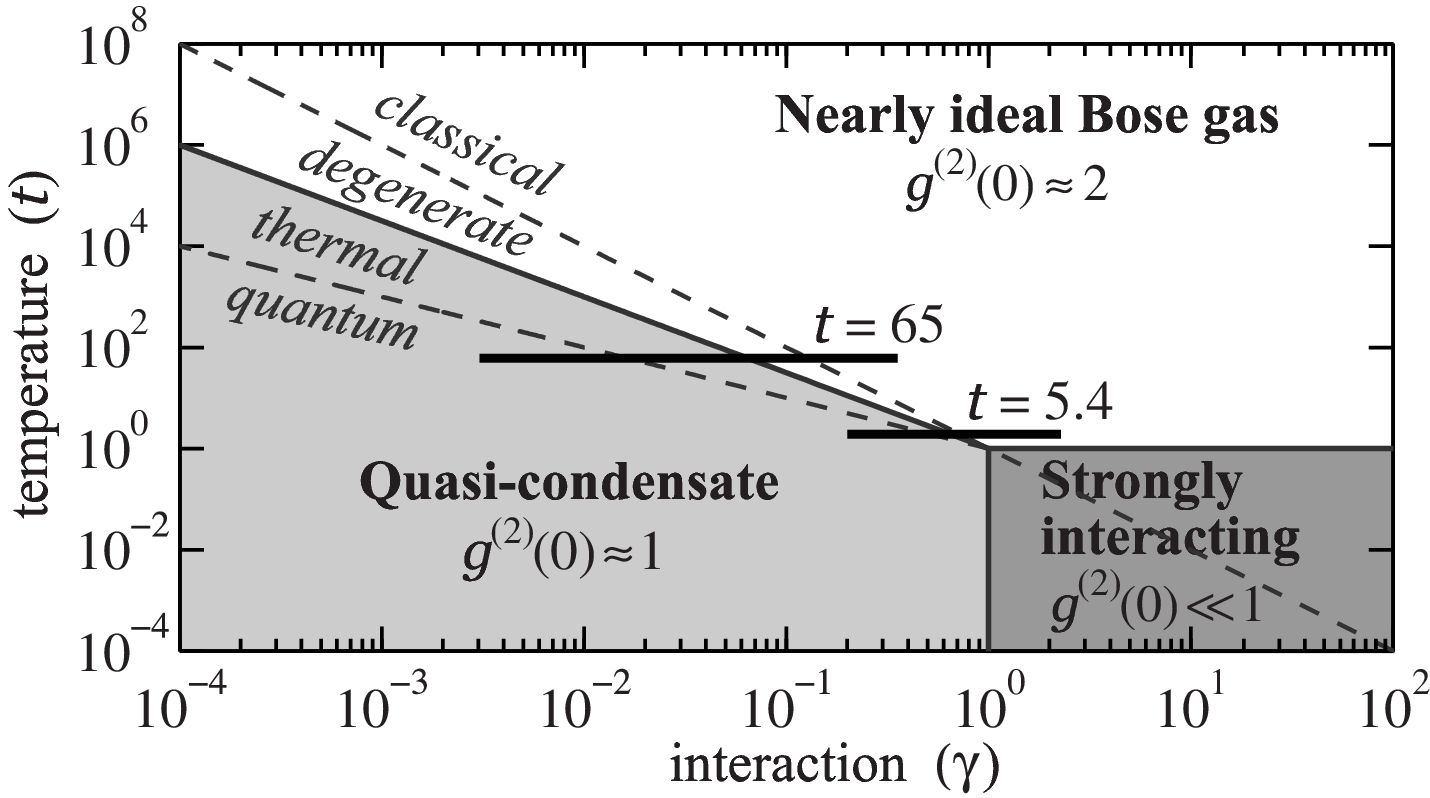} 
\caption{Phase diagram in the interaction-temperature
parameter space of a repulsive
uniform 1D Bose gas 
\cite{Kheruntsyan05}.
The values of the local two-body correlation $g^{(2)}(0)$ are indicated 
for the three main  
regimes (white and grey areas). 
The two horizontal lines show the parameters explored in this paper.
}
\label{fig.figphasediag}
\end{figure}

In this work,  we observe for the first time sub-Poissonian atom 
number fluctuations in 
small slices of a \textit{single}
one-dimensional (1D) Bose gas with repulsive interactions, 
where each slice approximates a uniform system.
 Taking advantage of the  long scale density variation due 
to a weak longitudinal confinement, we 
monitor -- at a given temperature -- the atom number fluctuations 
in each slice
as a function of
the local density.
For a weakly interacting gas, the
measured fluctuations are super-Poissonian at low densities, and they 
become sub-Poissonian
as the density is increased and the gas enters
the quantum quasi-condensate sub-regime that is 
dominated by \textit{quantum} rather than thermal fluctuations 
(see Fig.~\ref{fig.figphasediag}, with the interaction and temperature parameters, $\gamma$ and $t$, defined below).
When the strength of interactions is increased, the fluctuations are no longer 
super-Poissonian at low densities and remain 
sub-Poissonian at high densities.
The absence of super-Poissonian behavior implies 
that the gas enters the strongly 
interacting regime where the repulsive interactions 
between bosonic atoms mimic fermionic Pauli blocking,
and the quantities involving only densities
are those of an ideal Fermi gas.
Our results in all regimes are in good agreement with the exact 
Yang-Yang thermodynamic solution for the uniform 1D Bose gas 
with contact interactions~\cite{YangYang69}.

We recall that the thermodynamics of a uniform 1D Bose gas can be 
characterized by the dimensionless interaction and temperature parameters, 
$\gamma\!=\!mg/\hbar^2 n$ and  $t\!=\!2\hbar^2k_BT/mg^2$~\cite{Kheruntsyan05}, 
where $T$ is the temperature,
$n$ the 1D density, $g\simeq 2\hbar \omega_{\perp }a$ is the coupling constant, $a$ is the $s$-wave scattering length, 
and $\omega_{\perp}$ is the frequency of the transverse harmonic 
confining potential.
Figure~\ref{fig.figphasediag}
shows the  different regimes of the gas, 
characterized by the behavior of the 
two-body correlation function $g^{(2)}$ and
separated by 
smooth crossovers~\cite{Kheruntsyan05}.
Of particular relevance to the present work are the quantum quasi-condensate 
sub-regime where $g^{(2)}(0)\lesssim 1$~\cite{Kheruntsyan05},
and the strongly interacting regime where 
$g^{(2)}(0)\ll 1$~\cite{Kheruntsyan05,Weiss2005}.
The two different situations studied in this work 
are shown in Fig.~\ref{fig.figphasediag} 
by two horizontal lines at different values of $t$.

The experiments are performed using $^{87}$Rb atoms ($a\!=\!5.3$~nm) 
confined in a magnetic trap 
realised by current-carrying microwires on an atom chip. 
For the data at $t=65$, as in Ref.~\cite{Armijo10_2}, we use 
an H-shaped structure to
realise a very elongated harmonic trap at $\sim\!\!100~\mu$m 
away from the wires, with 
the longitudinal frequency $\omega_{\parallel}/2\pi\!=\!5.5$~Hz and 
$\omega_{\perp}/2\pi\!=\!3.3$~kHz.
 Using rf evaporation we produce clouds of $\sim\!\!3000$ atoms in
thermal equilibrium at $T\!=\!16.5~$nK, corresponding to $t\!=\!65$.
 We extract the longitudinal density profile from \textit{in situ} absorption
images as detailed in~\cite{ArmijoSkew}. The local atom number 
fluctuations in the image pixels, whose 
length in the object plane is $\Delta=4.5~\mu$m, are 
measured by repeating the same experiment 
hundreds of times and performing statistical 
analysis of the density profiles \cite{Armijo10_2}. 
For each profile and pixel, we record the 
atom number fluctuation $\delta N=N-\langle N\rangle$, 
where $\langle N\rangle=n\Delta$ is the mean atom number.
The results are binned 
according to $\langle N\rangle$ and for each 
bin we compute  the variance $\langle \delta N^2\rangle$.
The contribution of optical shot noise to $\langle \delta N^2\rangle$
is subtracted.

\begin{figure}
\includegraphics[width=6.5cm]{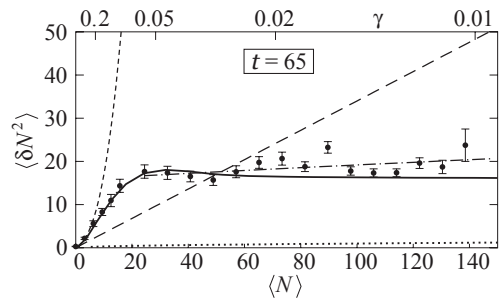} 
\caption{Variance of the atom number fluctuations in a weakly interacting gas, for $t=65$.
The measured data are shown as circles together with statistical uncertainties. 
Predictions from Eq.~(\ref{eq.thermo}) and different thermodynamic models are shown
as solid 
(Yang-Yang), short-dashed 
(ideal Bose gas)
and dash-dotted (quasi-condensate) lines. The long-dashed line
is the Poissonian limit and the low lying dotted line is the
contribution from quantum fluctuations.
Here $\omega_\perp/2\pi\!=\!3.3$~kHz, $\omega_\parallel/2\pi\!=\!5.5$~Hz,
$T\!=\!16$~nK~($k_BT=\!0.1\hbar\omega_\perp$), and 
$\kappa=0.34$.
}
\label{fig.varNchaud}
\end{figure}

Figure~\ref{fig.varNchaud} shows the measured variance $\langle \delta N^2\rangle$ versus $\langle N\rangle$. 
Since $l_c\ll\!\max\{\Delta,d\}\!\ll \!L$ in our experiment, where $L\!\simeq \!50$~$\mu$m is the cloud 
rms length, $d$ is the imaging resolution,
and $l_c$ is the correlation length of density 
fluctuations \cite{Kheruntsyan05,Deuar2009}, the local density 
approximation is expected to correctly describe both the average density profile and the 
fluctuations~\cite{Kheruntsyan05,Armijo10_2}. 
Accordingly, $\langle \delta N^2\rangle$ is expected to follow the 
thermodynamic prediction~\cite{Armijo10_2,ArmijoSkew} 
\begin{equation}
\langle \delta N^2\rangle =\kappa k_B T\Delta (\partial n/\partial \mu)_{T},
\label{eq.thermo}
\end{equation}
where $n(\mu,T)$ is the linear density of a homogeneous gas,
 and $\mu$ is the chemical potential.
The reduction factor $\kappa$ accounts for the finite resolution of the 
imaging system; it is determined from the measured
correlation between adjacent pixels~\cite{ArmijoSkew}, from which we deduce the rms
width of the imaging impulse response function $\cal A$ (assumed
to be Gaussian), and find that $d=3.5$~$\mu$m and $\kappa=0.34$.

The thermodynamic predictions for an ideal Bose gas and a quasi-condensate 
are shown in Fig.~\ref{fig.varNchaud}. In the quasi-condensate regime 
we use the equation of state $\mu=\hbar\omega_\perp(\sqrt{1+4na}-1)$~\cite{Fuchs03}.
The temperature is obtained by fitting 
the quasi-condensate prediction to the measured fluctuations 
at high densities. 
Usual features of the quasi-condensation
transition are seen~\cite{Armijo10_2}: at low density, the gas lies within
the ideal gas regime where, for degenerate gases,  
bosonic bunching raises the fluctuations well above the Poissonian limit;
at high density the gas lies in the quasi-condensate regime 
where interactions level off the density fluctuations.
Within the quasi-condensate regime, the fluctuations go from super-Poissonian
to sub-Poissonian, with $\langle \delta N^2\rangle/\kappa\langle N\rangle$ 
going from $2$ to $0.44$.
Using the approximate 1D expression $\mu\!=\!gn$,~Eq.~(\ref{eq.thermo}) shows that 
the transition 
from super- to sub-Poissonian behavior
occurs at  $k_BT\!\simeq\!gn$, which is the boundary between the thermal 
and quantum quasi-condensate regimes~\cite{Kheruntsyan05,Deuar2009}.
The fluctuations in the whole explored density domain are in good agreement
with  the exact 1D Yang-Yang predictions.  The small 
discrepancy at ~high densities~between the Yang-Yang and the quasi-condensate 
models is due to the transverse swelling of the cloud 
~\cite{ArmijoSkew,Armijo10_2}.
In the following, we neglect this  3D effect
and perform a purely 1D analysis.

Going beyond the thermodynamic relation~(\ref{eq.thermo}), 
the variance $\langle \delta N^2\rangle$ in a pixel is given by
\begin{equation}
\langle \delta N^2\rangle =\!\!\int \!\! d^{4}{\cal {Z}}~\langle \delta n(z)\delta n(z')\rangle \;{\cal A}(z-Z){\cal A}(z'-Z'),
\label{eq.fluctu}
\end{equation}
where $\int \!d^4{\cal {Z}}\equiv\!\!\int_0^\Delta \!\!dZ\!\!\int_0^\Delta \!\!dZ'\!\!\int_{-\infty}^{\infty}\!\!dz\!\!\int_{-\infty}^{\infty}\!\!dz'$,
$\delta n(z)=n(z)-\langle n(z) \rangle$ is the density fluctuation, and $\int_{-\infty}^{+\infty}\!dZ{\cal A}(Z)\!=\!1$.
Isolating the one- and two-body terms, one has
\begin{equation}
\langle \delta n(z)\delta n(z')\rangle=n\delta(z-z')
+n^2 [g^{(2)}(z-z')-1].
\label{eq.deltan}
\end{equation}
The first term, when substituted into Eq.~(\ref{eq.fluctu}), accounts for Poissonian 
level of fluctuations, $\kappa \langle N\rangle$.  Therefore,
the measured \textit{sub}-Poissonian
fluctuations in Fig.~\ref{fig.varNchaud}
imply that 
$g^{(2)}(z-z^{\prime})-1<0$.
Such 
anti-bunching stems from quantum fluctuations. Indeed, 
within the 
Bogoliubov approximation, valid
 for quasi-condensates,
 one has~\cite{Deuar2009}
 \begin{equation}
g^{(2)}(z\!-\!z')-1 \!=\int \frac{dk}{2\pi n} \left[2n_k f_k - (1-f_k)\right]e^{ik(z-z')},
\label{eq.g2Bogo}
\end{equation}
where $f_k=1/\sqrt{1+4/(l_\xi k)^2}$ 
and $n_k=1/\left(e^{\epsilon_k/k_BT}-1\right)$ is the thermal occupation 
of the Bogoliubov collective mode of wavenumber $k$ and energy 
$\epsilon_k=\hbar^2k^2/(2m\sqrt{1+4/(l_\xi k)^2})$, with 
$l_\xi=\hbar/\sqrt{mgn}$ being
the healing length.
The first term in the rhs of Eq.~(\ref{eq.g2Bogo}) 
which accounts for thermal fluctuations is positive, whereas the second term 
which is the contribution of quantum (i.e., zero temperature)
fluctuations is negative~\cite{Deuar2009}.
Therefore, the negativity of $g^{(2)}(z-z')-1$ 
implies that the quantum fluctuations give a larger contribution 
to $g^{(2)}(z-z^{\prime})-1$ than the thermal ones.

It should be emphasised, however, that the quantity we measure is 
$\langle \delta N^2\rangle $, and as we show below, for our large
values of $\Delta$ and $d$ it is still dominated 
by thermal 
(rather than quantum) 
fluctuations. 
This is because
the contribution to $\langle \delta N^2\rangle $ of the one-body 
term
almost cancels out the contribution 
of the zero-temperature two-body term.
Indeed, the contribution of quantum fluctuations to 
$\langle \delta N^2\rangle$, calculated using Eqs.~(\ref{eq.fluctu}),~(\ref{eq.deltan}), and (\ref{eq.g2Bogo}), 
is
\begin{equation}
\langle \delta N^2\rangle_{T=0}=
\frac{\langle N \rangle}{\Delta\pi}\int_{-\infty}^{\infty} 
dk f_k\frac{1-\cos(k\Delta)}{k^2} e^{-k^2d^2}.
\label{eq.fluctuquant}
\end{equation}
Since $f_k \propto kl_\xi$ when $kl_\xi\ll 1$, we find that 
for $\Delta\gg l_\xi,d$, $\langle \delta N^2\rangle_{T=0}$ scales 
as $n l_\xi \ln(\Delta/l_\xi)$. 
On the other hand, the thermal contribution 
given by Eq.~(\ref{eq.thermo}), scales as 
$\Delta T/g$. Therefore, the quantum 
contribution becomes negligible  
as $\Delta\rightarrow \infty$, 
and the thermodynamic prediction of Eq.~(\ref{eq.thermo}) is 
recovered~\cite{Klawunn2011}.
For our parameters, the contribution of Eq.~(\ref{eq.fluctuquant})  to
$\langle \delta N^2\rangle$ is 
shown as a dotted
line in Fig.~\ref{fig.varNchaud}. 

\begin{figure}
\includegraphics[width=8.5cm]{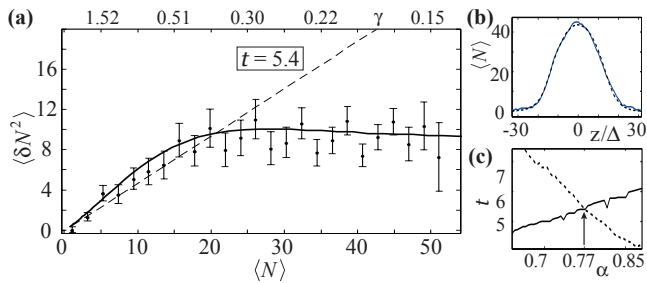} 
\caption{(a) Variance $\langle \delta N^2\rangle$ close to the strongly 
interacting regime, for $t=5.4$. Different curves 
are as in Fig.~\ref{fig.varNchaud}, but for  
$\omega_{\perp}/2\pi\!=\!18.8$~kHz,  
$\omega_{\parallel}\!=\!7.5$~Hz, $T\!=\!40$~nK~($k_BT\!=\!0.044\hbar\omega_\perp$),
and $\kappa\!=\!0.47$. 
(b) Average density profile (solid line) together with the Yang-Yang prediction (dashes). 
(c) The value of $t$ obtained from fits to the density profile (dotted line) 
and atom number fluctuations (solid line) for different $\alpha$ (see text).
}
\label{fig.varNfroid}
\end{figure}

In  weakly interacting gases, the atom number 
fluctuations take super-Poissonian values 
in the degenerate ideal gas and thermal quasi-condensate regimes,
$\langle \delta N^2\rangle/\langle N\rangle$ reaching its maximum  at 
the quasi-condensate transition where it  scales as
 $t^{1/3}$ \cite{Armijo10_2}.
When $t$ is decreased, the super-Poissonian zone is 
expected to merge towards the Poissonian 
limit and it vanishes when the gas enters the strongly interacting regime.
This trend is exactly what we observe in 
Fig.~\ref{fig.varNfroid}(a), for $t\!=\!5.4$:
at large densities, we 
see suppression of $\langle \delta N^2\rangle$ below the Poissonian level
but, most importantly, we no longer observe super-Poissonian fluctuations at lower densities
($\langle \delta N^2\rangle /\kappa \langle N\rangle\!<\!1.3$ within the 
experimental resolution) \footnote{
Fluctuations are, however, still much larger than those
of a Fermi gas for our (not very small) value 
of $t$.}.
Interestingly, no simple 
analytic 
theory is applicable to this 
crossover region, and the only reliable 
prediction here is the exact 
Yang-Yang 
thermodynamic 
solution [solid line in Fig.~\ref{fig.varNfroid}(a)].

\begin{figure}[t]
\includegraphics[width=8.4cm]{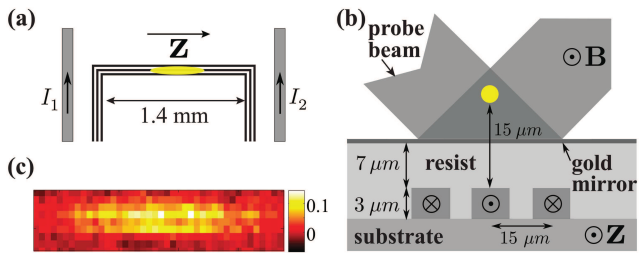} 
\caption{(color online) (a) Wire schematic of the atom chip: three gold wires 
along $\mathbf{Z}$ carry an ac
current and produce a tight transverse confining potential.
 The longitudinal
confinement is realized with dc currents $I_1$ and $I_2$. 
(b) The wires are buried under a layer of resist, 
which ensures electrical insulation and surface planarization. 
The resist is covered with  
200~nm thick gold  mirror that reflects the probe beam. 
The atoms are  $15\!$~$\mu$m away
from the wires and see the interference pattern produced by the 
 probe and the reflected beam. (c)
Typical optical-density image of a gas of $10^3$ atoms.
}
\label{fig.modulatedguide}
\end{figure}

We now describe  the experimental techniques that 
allowed us to
increase significantly $\omega_\perp$
in order to reach $t=5.4$.
Keeping a reasonable heat dissipation in the wires, 
increasing $\omega_\perp$ requires bringing the atomic cloud closer to the chip.
However, using dc micro-wire currents, one would observe  
fragmentation of the cloud 
due to wire imperfections and hence longitudinal roughness of the 
potential~\cite{Esteve2004}.
To circumvent this problem, we use the modulation techniques 
developed in~\cite{Trebbia2007,Bouchoule2008}. The atom chip schematic is 
shown in Fig.~\ref{fig.modulatedguide}. 
The transverse confinement is realized by three wires, carrying
the same ac current modulated at $200$~kHz,
and a longitudinal homogeneous 
dc magnetic field of $\sim\!1.8$~G realized by external coils. 
The  modulation is fast enough so that the atoms experience the
 time-averaged potential, transversely harmonic.
Monitoring dipole oscillations we measure 
$\omega_\perp/2\pi$ varying from $2$ 
to $25$~kHz, for ac current amplitude varying from $40$ to $200$~mA.
 The longitudinal confinement, with $\omega_z/2\pi$  varying   
from $5$ to $12$ Hz, is realised by
 wires perpendicular 
to the $z$-direction, carrying
 dc currents of a few tens of mA. After a 
first rf evaporation stage in a dc trap 
we load $6\times10^4$ atoms at a few $\mu$K in the ac trap where we 
perform further rf evaporation at $\omega_{\perp}/2\pi\simeq 2$~kHz 
and $\omega_{\parallel}/2\pi\simeq 12$~Hz. Next we lower the 
longitudinal trapping frequency to about $7$~Hz and then ramp up the 
transverse frequency to $18.8$~kHz 
in $600$ ms keeping the rf evaporation on 
during this compression. After ramping the rf power 
down in $100$ ms and letting 
the cloud to thermalize for $150$ ms, we switch off the wire currents and image 
the atomic cloud after $50$~$\mu$s   
with a $60$~$\mu$s long resonant probe pulse.
The probe 
is circularly polarised and its intensity, chosen to 
 optimise the signal to noise ratio, is about $0.2\, I_{\mathrm{sat}}$,
where $I_{\mathrm{sat}} = 1.67$~mW is the saturation intensity of the 
D$_2$ line.
Finally, we get 
the longitudinal profile of the cloud by 
summing over the transverse pixels.
We typically obtain clouds at $t\simeq 1-6$.
Taking a few hundreds 
of images under the same conditions, we  
measure $\langle \delta N^2\rangle$ 
the same way as for the results of Fig.~\ref{fig.varNchaud}.

One crucial point to correctly determine the longitudinal profile 
is the knowledge of the absorption cross section $\sigma$. 
In our setup, 
atoms sit in an interference pattern during the imaging pulse [see Fig.~\ref{fig.modulatedguide}(b)] 
and are subjected to a magnetic field so that 
the determination of $\sigma$ is not simple.
Following~\cite{Reinaudi}, we assume $\sigma = \alpha \sigma_\mathrm{o}/\left(1+\alpha I/I_{\mathrm{sat}}\right)$, where $I$ is the intensity of 
the probe beam, $\sigma_\mathrm{o}=3\lambda^2/2\pi$ is the resonant 
cross section 
of the transition 
$|F=2, m_F=2\rangle \rightarrow  |F'=3, m_F'=3\rangle$, 
and $\alpha$ is a numerical factor. Solving the 
optical Bloch equations (OBE) for our probe intensity and duration, 
we find that such a law is valid, and we obtain $\alpha = 0.75$. 
In this calculation, we averaged $\alpha$ over the distance to the chip, 
which is expected to be valid as atoms diffuse over a rms width of 
about $1~\mu m$ 
during the imaging pulse, which is larger than the interference lattice period.
The factor $\alpha$ can be also deduced from the mean density 
profile and/or the atom number fluctuations using the thermodynamic 
Yang-Yang predictions. 
Fitting both $\alpha$ and $t$ to either the mean profile or the 
fluctuations leads to strongly correlated values of $\alpha$ and $t$ but with large 
uncertainty in $\alpha$. 
Combining both pieces of information, however, enables a precise 
determination of $\alpha$.
More specifically, using the Yang-Yang theory, we extract $t_p$ and $t_f$ from fits
to the mean profile and the fluctuations, respectively, 
for various values of $\alpha$ [see Fig.~\ref{fig.varNfroid}(b) and (c)].
The intersection $t_p=t_f$ gives the correct value of $\alpha$.
 We find  $\alpha=0.77$, in good 
agreement with the OBE calculation. The corresponding value of $t$ is $5.4$ and hence  $T=40$~nK.

In summary, we have realised for the first time a \textit{single} 1D Bose gas close to the strongly
interacting regime. In contrast to realisations of 
arrays of multiple 1D gases in 2D optical lattices \cite{Phillips-2004,Weiss2005}, 
our experiments have allowed us to 
perform atom number fluctuation measurements in small 
slices of the gas, not possible with multiple 1D gases.
In the weakly interacting regime, we 
reached the \textit{quantum} quasi-condensate regime (where  $k_BT\!<\!\mu\!=\!gn$) in a strictly 1D situation with $k_BT\!\ll\!\hbar\omega_\perp$. 
Although the two-body correlation function $g^{(2)}$ is
dominated by quantum fluctuations in this regime, we have 
shown that the variance $\langle \delta N^2\rangle$ is 
still dominated by thermal excitations. 
To resolve quantum fluctuations one would need to access wavelengths 
smaller than 
the phonon thermal wavelength $\hbar^2/mk_BT l_\xi$ 
\cite{Klawunn2011}, which is in the submicron range for our parameters.
Our work opens up further opportunities in the study of 1D Bose gases, 
such as better understanding of the mechanisms of thermalisation and the 
role of three-body correlations.

\begin{acknowledgments}
The authors acknowledge support by the IFRAF Institute, the
ANR Grant No. ANR-08-BLAN-0165-03, the ARC Discovery Project Grant No. DP110101047, and the CoQuS Graduate school of the FWF and the Austro-French FWF-ANR Project I607.
\end{acknowledgments}


\end{document}